\documentclass{ws-ijmpa}
\usepackage[super,compress]{cite}
\usepackage{graphicx}
\graphicspath{{figvj/}}

\def\alp{ALPGEN}

\def\she{SHERPA}

\def\bhs{{\sc BlackHat}+SHERPA}
\def\meps{ME$\texttt{+}$PS}
\def\mg{MADGRAPH}
\def\mca{MC@NLO}
\def\mcaher{MC@NLO$\texttt{+}$HERWIG}

\def\pow{POWHEG}
\def\powpyt{POWHEG$\texttt{+}$PYTHIA}
\def\mepsatnlo{MEPS@NLO}
\def\jph{{\sc Jetphox}}

\def\ifb{\ensuremath{\rm fb^{-1}}}

\def\rts{\ensuremath{\sqrt{s}}}
\def\pp{\ensuremath{pp}}
\def\pap{\ensuremath{p\bar{p}}}
\def\tev{\ifmmode {\mathrm{\ Te\kern -0.1em V}}\else \textrm{Te\kern -0.1em V}\fi}%
\def\gev{\ifmmode {\mathrm{\ Ge\kern -0.1em V}}\else \textrm{Ge\kern -0.1em V}\fi}%
\def\mgeq{\ensuremath{\geq}}
\newcommand{\antikt}{anti-$k_{t}$}

\def\htj{\ensuremath{H_{\rm T}}}
\def\ptj{\ensuremath{p^{\rm jet}_{\rm T}}}
\def\ptje{\ensuremath{p^{\rm jet1}_{\rm T}}}
\def\ptjz{\ensuremath{p^{\rm jet2}_{\rm T}}}

\def\mjj{\ensuremath{m_{jj}}}

\def\antibar#1{\ensuremath{#1\bar{#1}}}
\def\ttbar{\antibar{t}}

\def\Zg{\ensuremath{Z}}

\def\Wln{\ensuremath{W\to\ell\nu}}

\def\Wmn{\ensuremath{W\to\mu\nu}}
\def\Zgll{\ensuremath{\Zg\,(\to\ell \ell)}}
\def\Vjets{\ensuremath{V\,\texttt{+}\,\mathrm{jets}}}
\def\Wjets{\ensuremath{W\,\texttt{+}\,\mathrm{jets}}}
\def\Zjets{\ensuremath{\Zg\,\texttt{+}\,\mathrm{jets}}}
\def\Gjets{\ensuremath{\gamma\,\texttt{+}\,\mathrm{jets}}}
\def\ttjets{\ensuremath{\ttbar\,\texttt{+}\,\mathrm{jets}}}
\def\Vj{\ensuremath{V\,\texttt{+}\mgeq\mathrm{1~jet}}}
\def\Vjj{\ensuremath{V\,\texttt{+}\mgeq\mathrm{2~jets}}}
\def\Zjj{\ensuremath{\Zg\,\texttt{+}\mgeq\mathrm{2~jets}}}

\def\Wlnjets{\ensuremath{\Wln\,\texttt{+}\,\mathrm{jets}}}
\def\Wmnjets{\ensuremath{\Wmn\,\texttt{+}\,\mathrm{jets}}}
\def\Zlljets{\ensuremath{\Zgll\,\texttt{+}\,\mathrm{jets}}}

\begin{document}
\markboth{U.~Blumenschein}{Jets in Association with Vector Bosons or Top}

%
\catchline{}{}{}{}{}
%


\title{Jet Production in Association with Vector Bosons or Top Quarks}

\author{Ulla Blumenschein}

\address{II Physikalisches Institut, University of Goettingen, Friedrich-Hund-Platz 1\\
Goettingen, 37077, Germany\\
ublumen@gwdg.de}

\maketitle

\begin{history}
\received{24 05 2015}
\revised{}
\end{history}

\begin{abstract}
The LHC experiments ATLAS and CMS have measured  \Vjets\ and \ttjets\ final states over a large energy range  in data collected between 2010 and 2012 at $\rts = 7\tev$ and $\rts = 8\tev$. The results have been compared to  pQCD calculations at NLO and have been used to validate novel Monte Carlo techniques.

\end{abstract}

\ccode{PACS numbers:}


\section{Introduction \label{sec:Intro}}

Processes with leptons or photons  in association with hadronic  jets  in the final state play an important role in the physics program of the 
LHC. They constitute in particular  a typical signature of the production of  top pairs and Higgs bosons. Moreover, 
an accumulation of these processes in certain extreme regions of phase space would be an indication
of new BSM physics. 
Important backgrounds in the selection of Higgs and BSM processes are the associated production of  gauge  bosons or top quarks with jets ($\ttjets$, $\Vjets$).
Typically,  the multiplicity and kinematics of jets  in  these events are exploited  to achieve a separation  from the signal, which makes
the  precise modelling of these quantities a high priority.

\Vjets\ and \ttjets\ processes also constitute  powerful tests of perturbative quantum chromodynamics (pQCD), with the mass of the 
bosons or the top quark providing a  well-defined scale.
Photons and leptonic decay products of massive bosons or top quarks allow to trigger the event independently of the  jet kinematics.  
On the theoretical side, the large progress in  automated calculation of loop amplitudes~\cite{blackhat} allows for next-to-leading order (NLO) predictions with respect to a  fixed order of up to five additional partons.  A second path pursues the combination of tree level matrix elements~\cite{alpgen,sherpa,madgraph} or even  NLO matrix elements~\cite{sherpav2} of different parton multiplicities matched to a parton shower  (\meps ,  \mepsatnlo ).
Since the dominant graphs in basic processes contain at least one gluon in the initial state, differential or double-differential measurements of angles and momenta or masses can be used to constrain the gluon parton distribution function (PDF).

The signal selections start with a standard selection  of inclusive $W$, $Z$, $\gamma$ or  \ttbar\ final states, extended by additional requirements on the presence of jets.
Measured yields are typically unfolded to the particle level using regularized or iterative methods~\cite{svd,dagostini} and compared with fixed-order 
calculations and generator  predictions. The latter are corrected by global k-factors  to the most precise available calculation for the inclusive process.
Jets are reconstructed with the \antikt\ algorithm~\cite{antikt}, with a distance parameter of $R=0.4$ in ATLAS and  $R=0.5$  in CMS.

\section{Measurements of the V+jets Cross Sections \label{sec:Vjets}}

\Vjets\ cross sections are typically measured in clean final states with electrons and muons.
Although the cross section for leptonically decaying $W$ bosons is  larger than for $Z$ bosons by an order of magnitude, measurements are published  for both processes.
Not only  is it of interest to probe the impact of the small mass difference between the bosons but in particular both processes complement each other in terms of uncertainties:  the \Wlnjets\ selection profits from the larger cross sections, whereas the \Zlljets\ final state can be selected with a higher purity.

Before the start of the LHC, \Vjets\ cross sections have been measured at  the Tevatron in \pap\ collisions at center-of-mass energies of $\rts = 1.96 \tev$~\cite {cdfzjets,cdfwjets,d0zjetsa,d0zjetsb}, where the production is dominated by quark annihilation. For moderate transverse momentum regimes, the results confirmed the NLO  calculations for \Vj\ and \Vjj ~\cite{z2jetsnlo}. The data was also compared with early releases of  \meps\ generators. 

In contrast,  \Vjets\ production in  \pp\ collisions is dominated  by the Compton process. The larger center-of-mass energies by the LHC enhance the  QCD radiation considerably and allow to explore large jet multiplicities and large momentum ranges  relevant for Higgs  and BSM physics.
ATLAS and CMS experiments complement each other, both in the selection of observables and  of the theory predictions.  Typical minimal jet transverse momenta are $\ptj \ge 30 \gev$,  considerably lower than in the di-jet or multi-jet measurements.

During the first years of LHC data taking, novel NLO techniques~\cite{blackhat}  resulted in a series of new fixed-order predictions  for the production of massive gauge bosons with up to five jets~\cite{bhwjets,bhzjets}. The LHC experiments have used the data collected in 2011 at  $\rts = 7\tev$, corresponding to an integrated luminosity of $4.6-5.0~\ifb$  in order to probe these new predictions~\cite {atlzjets11,atlwjets11,atlrjets11,atlzjets12,cmszjets11,cmswjets11}. In addition, the performance of LO and NLO \meps\  generators was studied.
The results confirm the validity of the fixed-order NLO predictions for vector bosons produced in association with up to five jets performed with the new  techniques.  In  general,  \meps\  generators show a  reasonable performance  in modelling inclusive and differential jet cross sections.
Figure~\ref{f:Vjets1}a shows exemplarily the \Wjets\ cross section with respect to the jet multiplicity measured by the CMS experiment~\cite{cmswjets11} compared with fixed-order NLO calculations and predictions from \meps\ generators. 
Figure~\ref{f:Vjets1}b shows the ratio of \Wjets\ and \Zjets\ cross sections as a function of the exclusive jet multiplicity~\cite{atlrjets11}.  The ratio is well modelled both by fixed-order calculations and  the generators. 
 
The good description is preserved for exclusive jet multiplicities, even after increasing the scale difference between the  leading jet and the  jets which are vetoed to define the exclusive final state or after selecting events with large di-jet masses and rapidity distance~\cite{atlzjets11}. 

The data at center-of-mass energies of $7\tev$ and $8\tev$ allows to explore the event topology in a kinematic range up to the \tev\ scale.  The list of observables includes jet transverse momenta and rapidity, angles between the final-state particles  and the di-jet mass. The measurements show in general a good performance of fixed-order  calculations and \meps\ generators but also reveal some deficiencies.  
As an example, Fig.~\ref{f:Vjets2}a shows the theory/data ratio of the \Wjets\ cross section measured by the ATLAS experiment as a function of the  scalar transverse momentum sum \htj\ of all final state objects~\cite{atlwjets11}. The fixed-order \Wjets\  calculation underestimates large values of \htj\  characterized by  larger parton  multiplicities. \meps\ generators prove to be better suited for modelling the  \htj\ observable  but  overestimate  the cross section for hard jets. The performance of \she\  is improved by including explicit NLO matrix elements (\mepsatnlo ). 
Figure~\ref{f:Vjets2}b shows exemplarily the \Zjj\ cross section normalized to the inclusive $Z$  cross section as a function of the dijet mass  compared with predictions from two generators~\cite{atlzjets12}. The distribution is well modelled at NLO by the generator \pow ~\cite{powheg}.


\begin{figure}
\centering
 \includegraphics[width=0.43\textwidth]{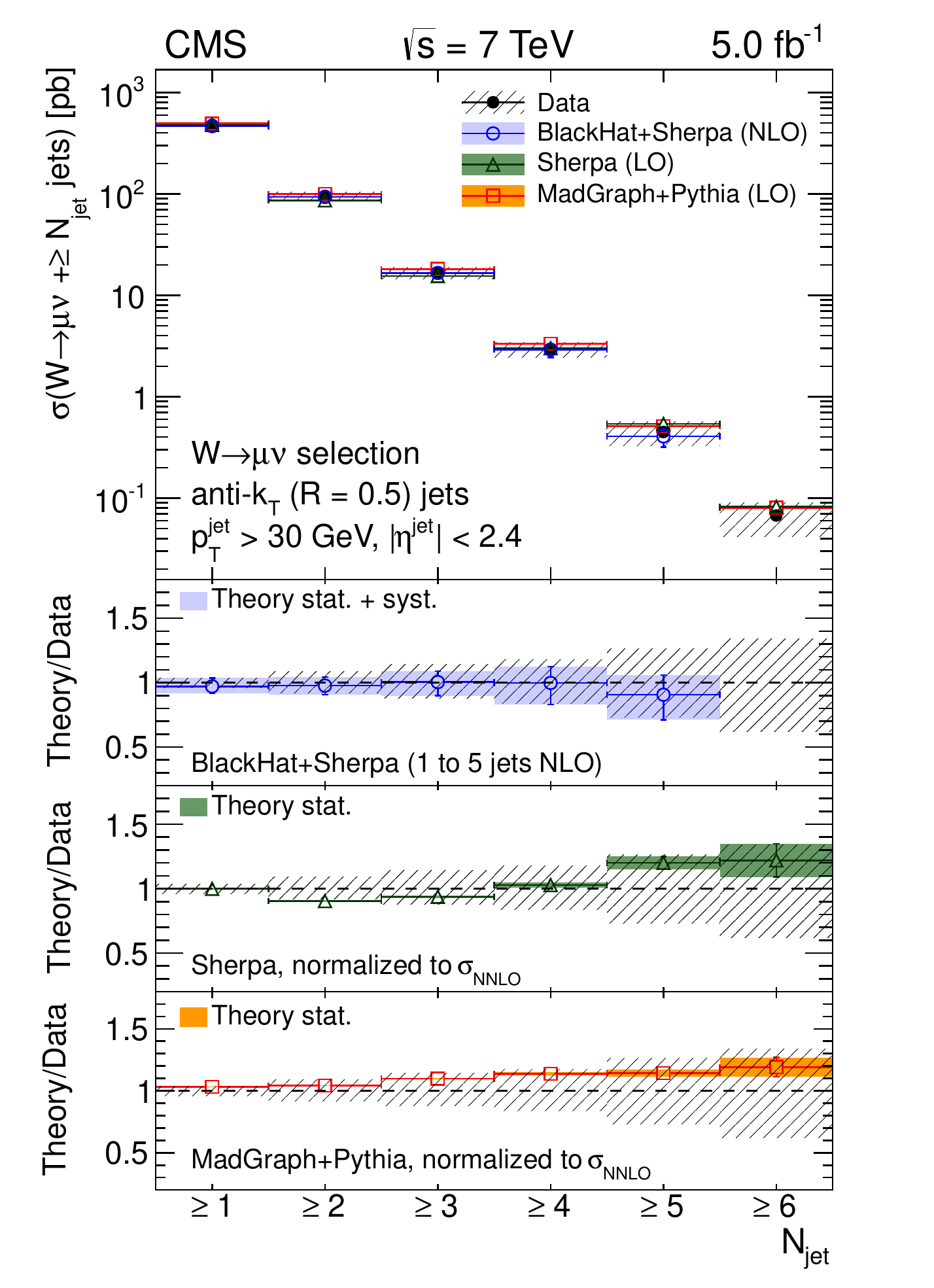}
 \includegraphics[width=0.45\textwidth]{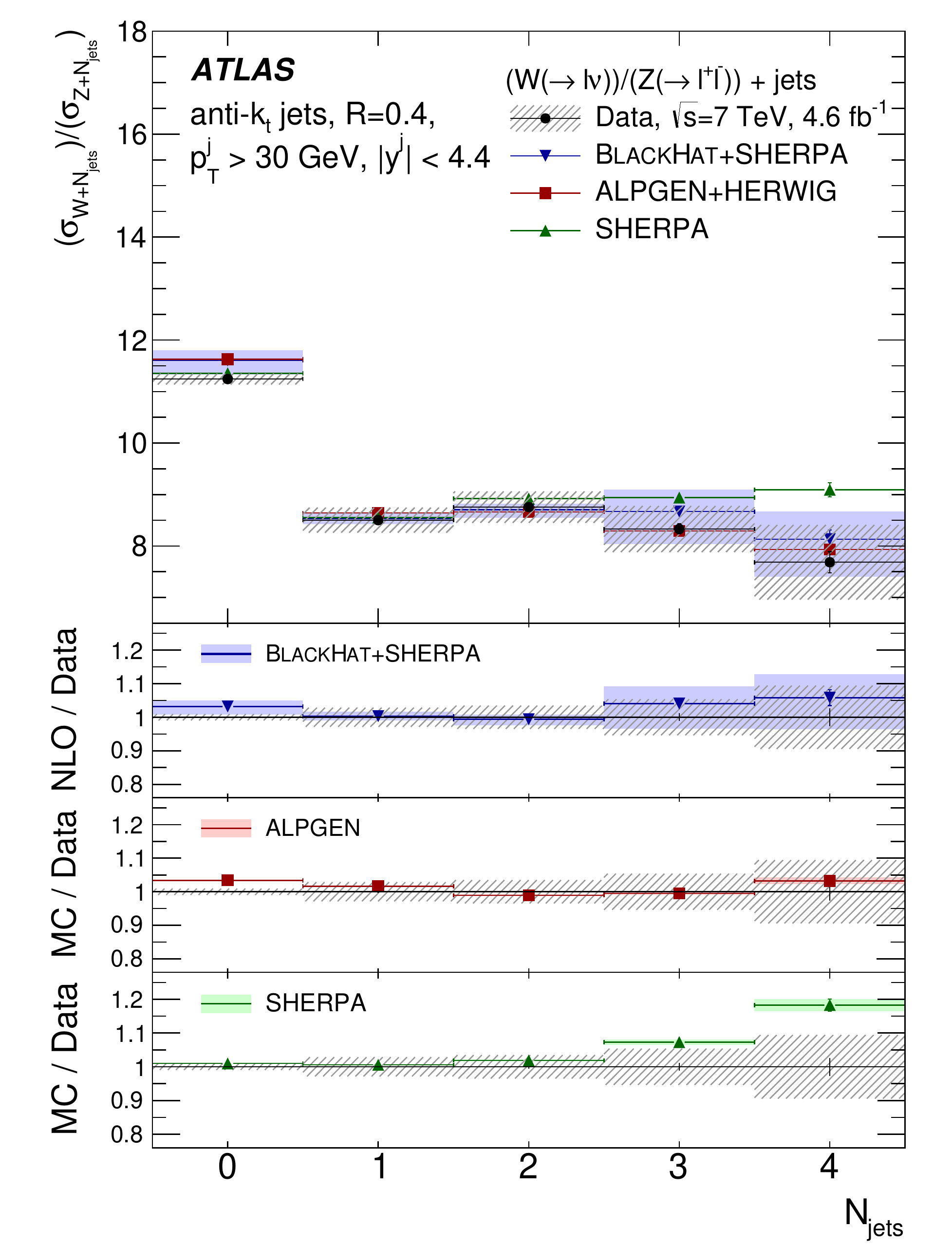}
 \caption{(a) \Wmnjets\  cross section with respect to the inclusive jet multiplicity measured by the CMS experiment~\cite{cmswjets11}.  (b)   Ratio of \Wjets\ and \Zjets\ cross sections as a function of the exclusive jet multiplicity measured by the ATLAS experiment~\cite{atlrjets11}. The data is compared with fixed order calculations by \bhs\ and with predictions from the \meps\ generators \she , \mg\  and \alp . \label{f:Vjets1}} 
\end{figure}

\section{Measurements of the Photon+Jets Cross Sections  \label{sec:Gjets}}

Isolated  photons  in association with jets provide an important 
probe of pQCD over a large energy range  in a comparatively clean final state.
The production of these events is dominated by the Compton process, such that they can be used in addition to the inclusive photon data to constrain the gluon PDF.
In addition, \Gjets\ constitutes a major background  to the Higgs$\to\gamma\gamma$ process and to several BSM topologies.

The LHC experiments have measured  \Gjets\ cross sections in the 2010 and in the 2011 data sets  at  $\rts = 7\tev$~\cite{atlgjets10a,atlgjets10b,cmsgjets11}. 
Typically, a photon isolation requirement is employed in order to enhance the purity of prompt direct photon production with respect to prompt photons  from
 fragmentation and  non-prompt  photons from hadron decays  in di-jet events. The published cross sections include the remaining  fragmentation component.
Differential cross sections are measured with respect to characteristic event quantities. In order to optimize the PDF sensitivity, double or triple differential cross sections are extracted  as a function of the photon transverse energy, the photon pseudorapidity and  the jet pseudorapidity. 
The unfolded data are compared with  fixed-order NLO pQCD calculations by the \jph\  programme~\cite{jetphox} and with predictions from \meps\ generators. Photon and jet minimum transverse momenta are typically in the range of 30-45~\gev .
Dominant systematic uncertainties arise from the determination of the background from non-prompt photons.


\begin{figure}
\centering
\includegraphics[width=0.43\textwidth]{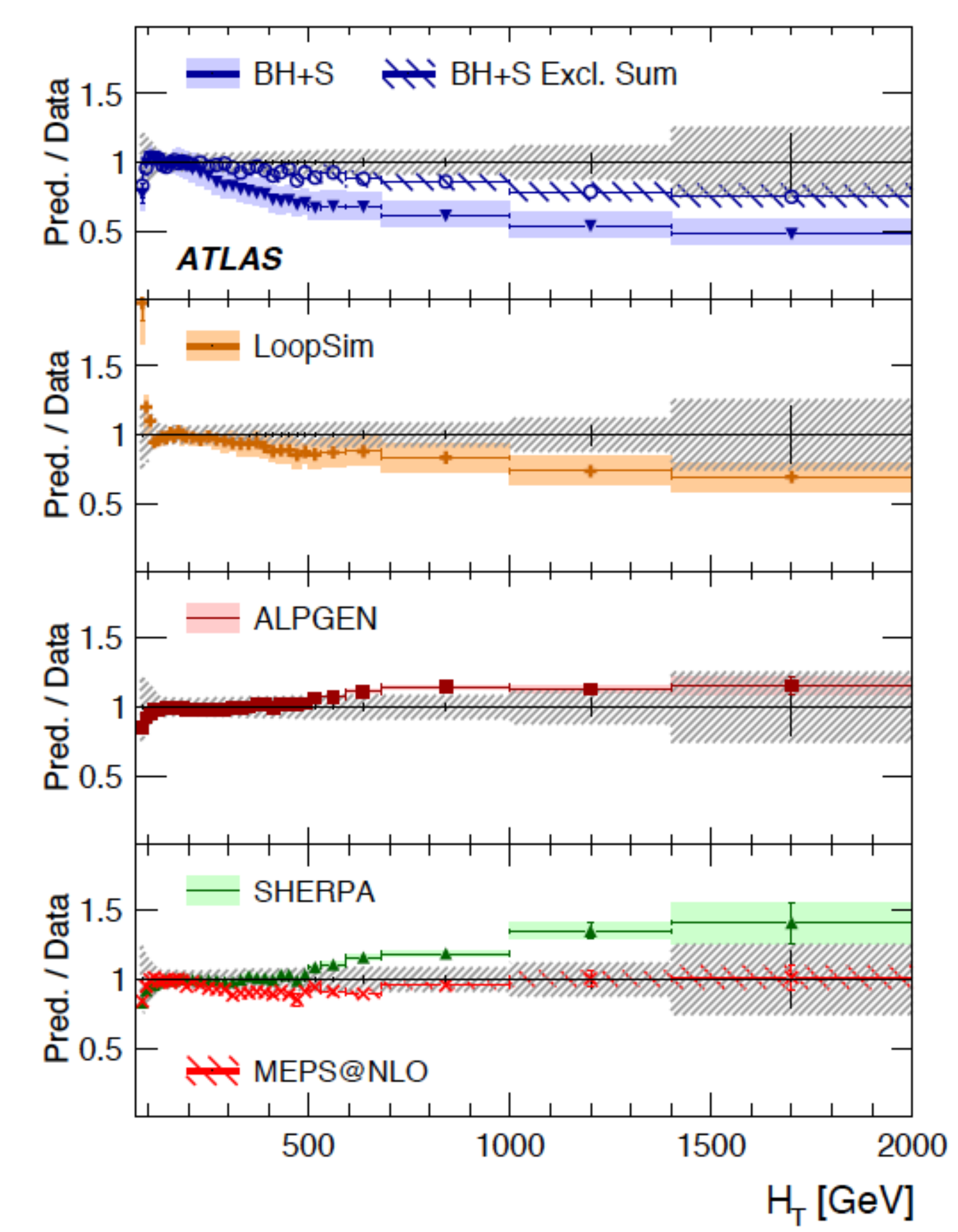}
\includegraphics[width=0.45\textwidth]{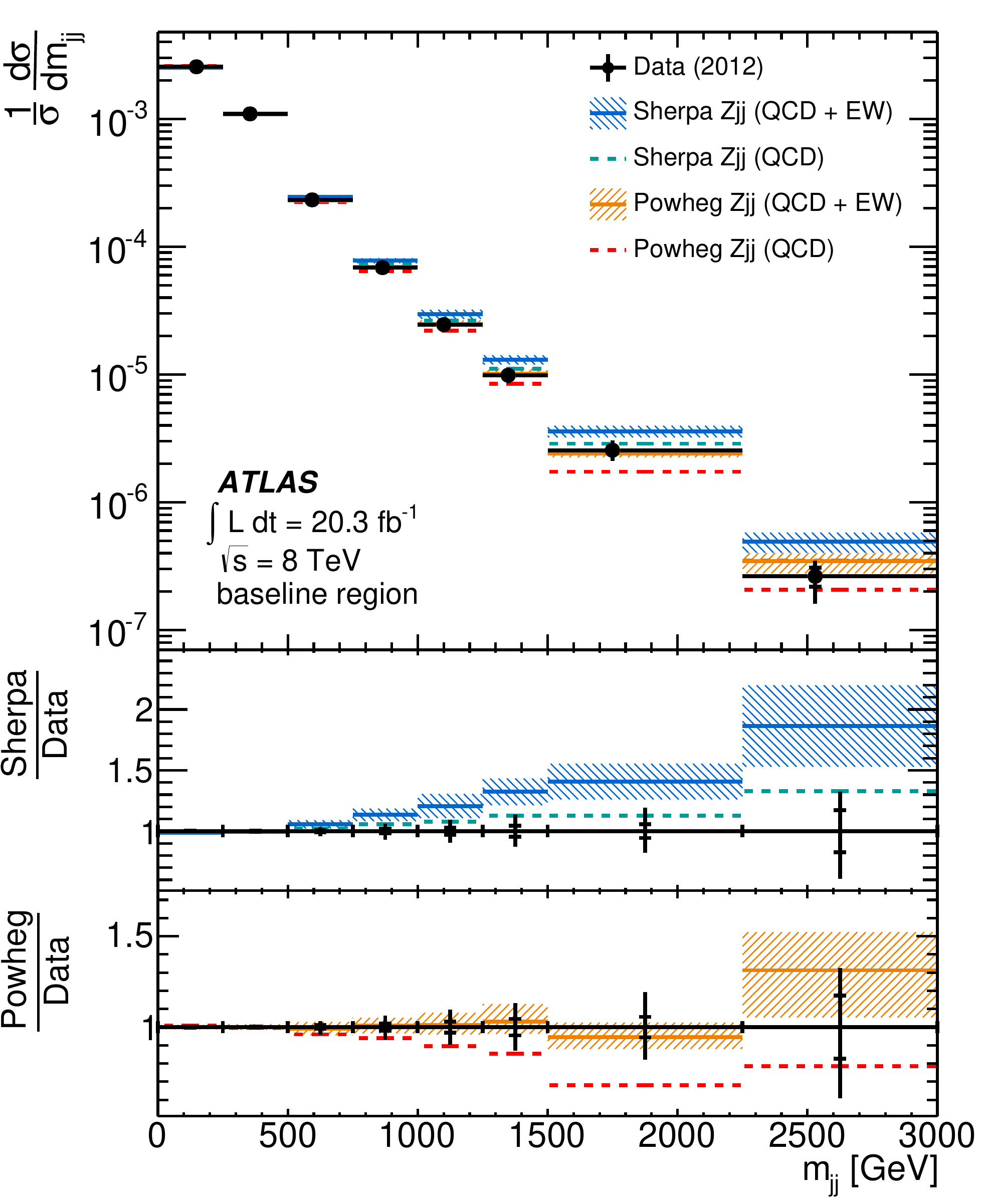}
\caption{(a) $\Wjets$ cross section  as a function of \htj\ in data
     collected with the ATLAS experiment at $\rts = 7\tev$~\cite{atlwjets11}. (b)  \Zjj\ cross section measured by the ATLAS experiment  normalized to the inclusive $Z$  cross section as a function of the dijet mass \mjj\ for  events with  $\ptje > 55\gev$ and $\ptjz > 45\gev$~\cite{atlzjets12}.  The data is compared with fixed order calculations  by \bhs\ and with predictions from the \meps\ generators \she\ and \alp\ and the generator \pow . \label{f:Vjets2}} 
\end{figure}

As an example, Fig.~\ref{f:Vjets5}a shows  an early  measurement performed by ATLAS  of the  scattering angle  $\theta^*$   in the photon-jet centre-of-mass frame in a region not distorted by the  kinematic preselection.  The measurement confirmes the validity of the NLO pQCD calculation by \jph\  and in particular the dominance of the Compton process in the isolated \Gjets\  sample. Figure~\ref{f:Vjets5}b shows part of the triple-differential cross section measurement performed by CMS  as a function of the photon transverse energy, the photon pseudorapidity and the jet pseudorapidity in  the  2011 data set. The double and triple differential LHC data confirms again the good performance of  the  pQCD calculations and of the predictions by the LO \meps\ generator \she . $\Gjets$ data collected  at  $\rts = 7\tev$ is  expected to constrain the gluon PDF by up to 20\%~\cite{photjetpdf}.

\begin{figure}
\centering
\includegraphics[width=0.46\textwidth]{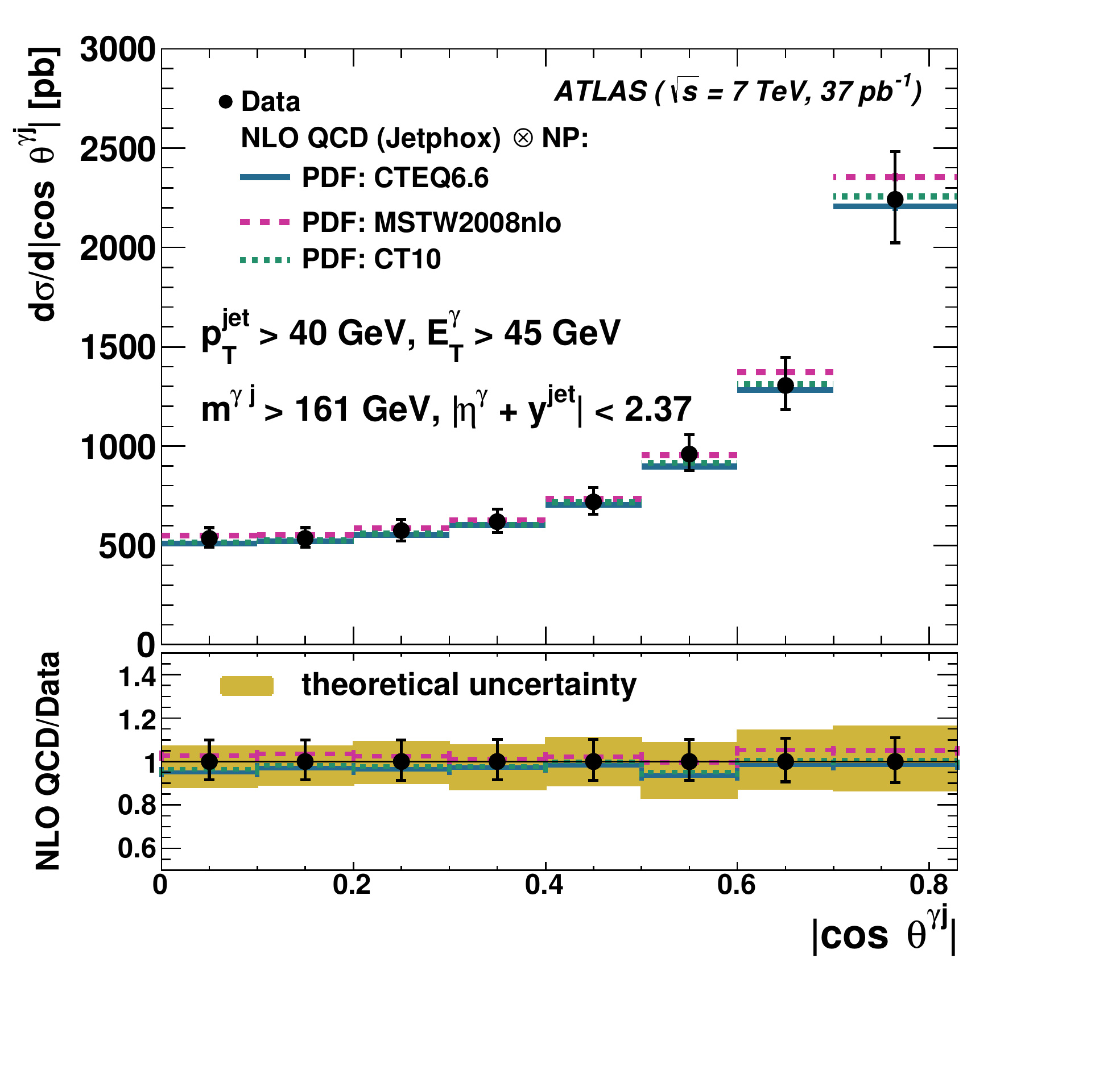}
\includegraphics[width=0.46\textwidth]{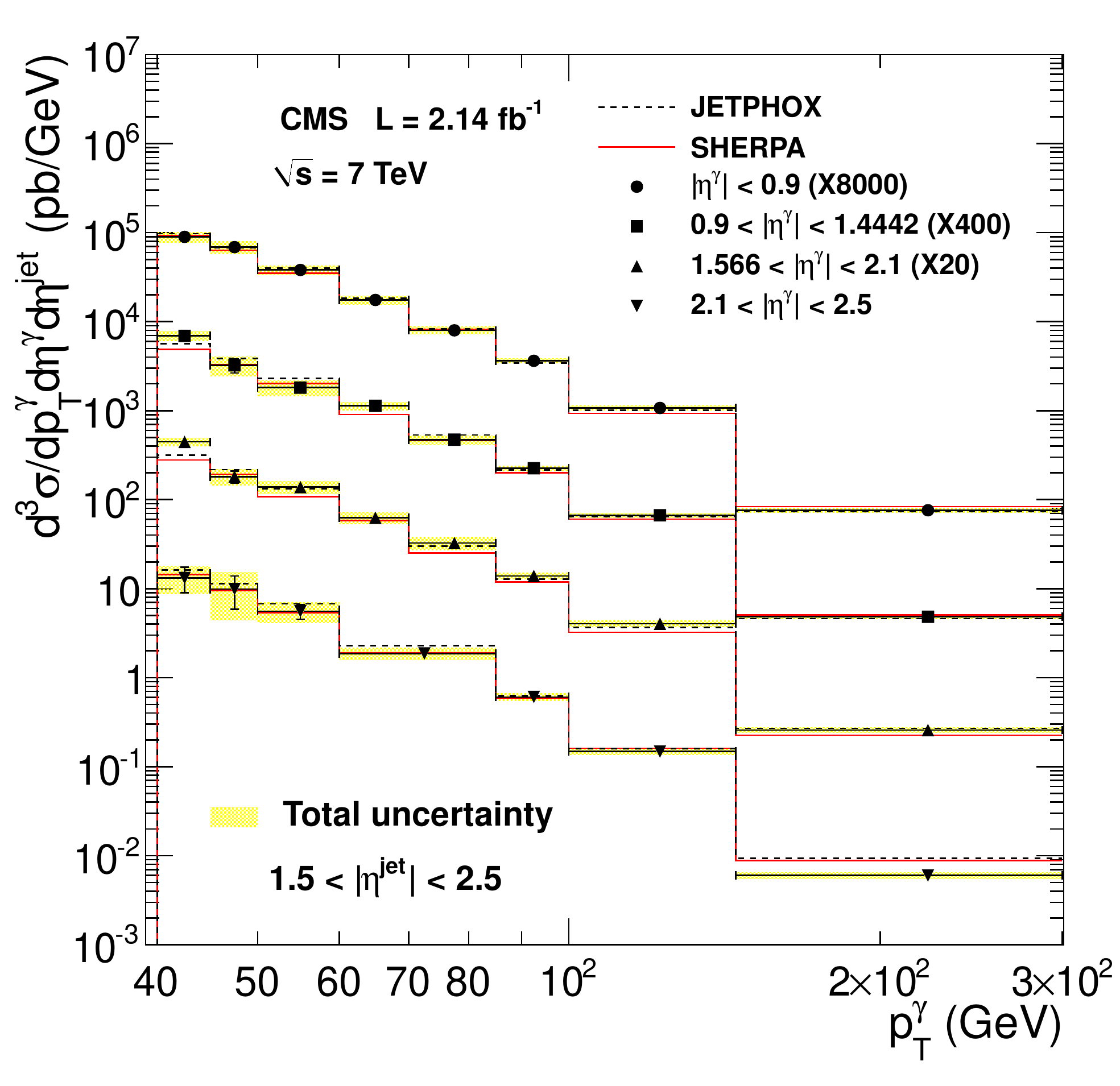}
\caption{(a) Differential cross section for  isolated $\Gjets$ with respect to $\rm cos\theta^*$ measured  by the ATLAS experiment~\cite{atlgjets10b}. (b) Part of the triple-differential $\Gjets$  cross section measurement performed by CMS~\cite{cmsgjets11}.  \label{f:Vjets5}}
\end{figure}

\section{Measurements of the \ttjets\ Cross Sections  \label{sec:TTjets}}

About half of the \ttbar\ events produced at the LHC are accompanied by additional hard hadronic jets which do not originate from the top quark decay. These 
events form part of the signal selection for  \ttbar\ precision measurements and constitute  an important background to  processes with multijet final states, in particular in the Higgs and BSM sector. They also serve as  an  important probe of pQCD with the top quark mass providing a large well-defined scale.

ATLAS and CMS have measured \ttjets\ cross sections in data collected in 2011 at $\rts = 7\tev$. Both experiments  select \ttbar\ events in the  di-lepton  and in the lepton+jets final states with two resp. four jets expected from the \ttbar\ decay itself. The dominant systematic uncertainty for large jet multiplicities arises from the jet energy scale. Cross sections are published for various minimum \ptj\  thresholds. The ATLAS collaboration measures cross sections as a function of the jet multiplicity and the jet transverse momenta. CMS publishes cross sections as a function of the jet multiplicity   normalized to the inclusive \ttbar\ cross section to cancel part of the systematic uncertainties.   Measurements are compared to predictions of \meps\ generators, where additional radiation of up to three jets is modelled by explicit matrix elements, and to NLO predictions from \pow\ and \mca ,  where the production of more than one additional jet is modelled by the parton shower.

 Figure~\ref{f:Vjets6}a shows exemplarily the \ttjets\ cross section as a function of the multiplicity of jets with $\ptj > 25\gev $ measured by the ATLAS experiment in the lepton+jets channel~\cite{atltjets11}.  
Figure~\ref{f:Vjets6}b shows  the corresponding normalized measurement in the dilepton channel, performed by the CMS experiment for jets with $\ptj > 30\gev $~\cite{cmstjets11}. The LHC results demonstrate  that \mcaher\ underestimates the radiation of more than one additional parton, while the \meps\ generators and \powpyt\ provide a reasonable description of  the additional jet activity for all \ptj\ thresholds.

\begin{figure}
\centering
\includegraphics[width=0.42\textwidth]{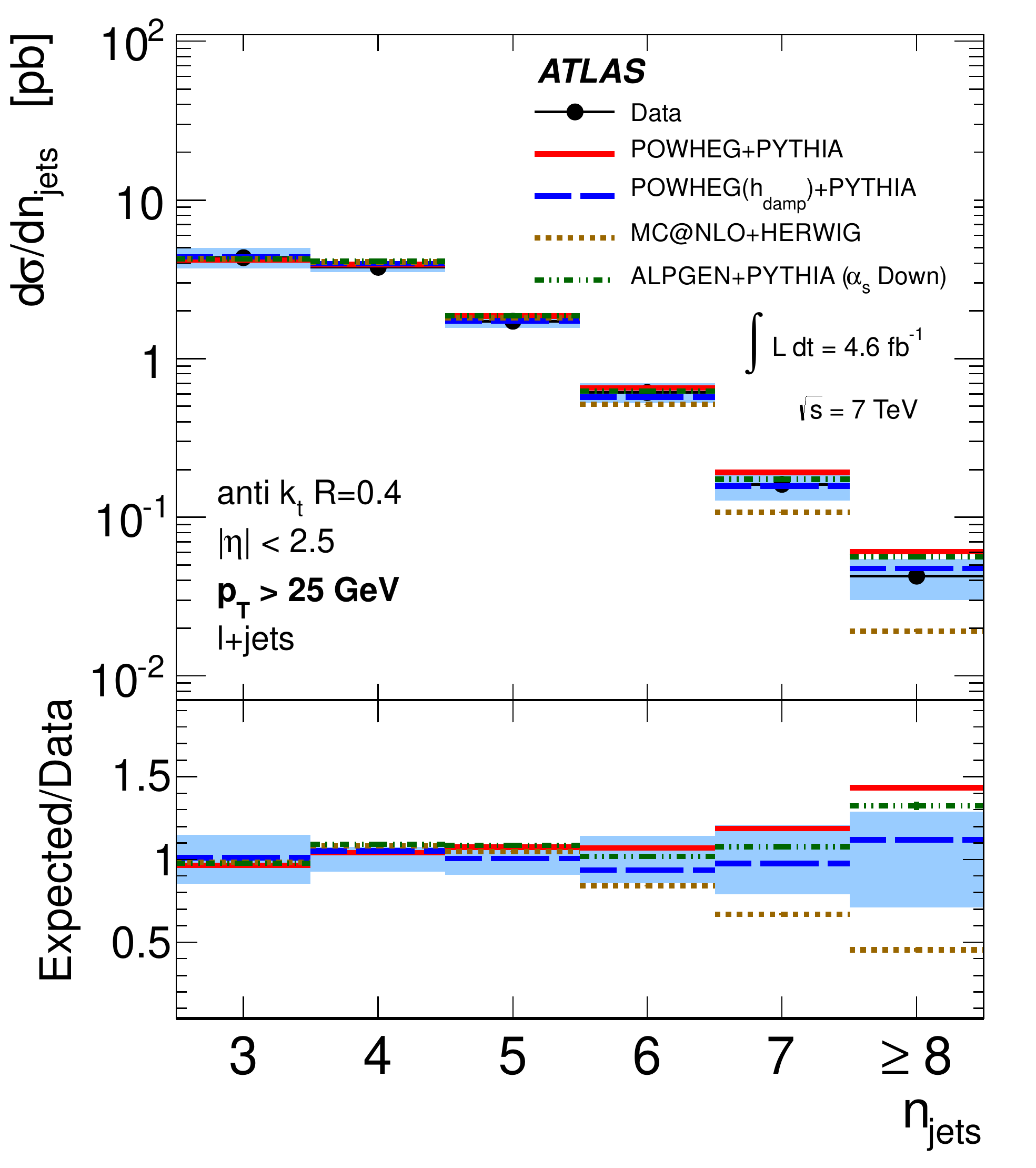}
\includegraphics[width=0.48\textwidth]{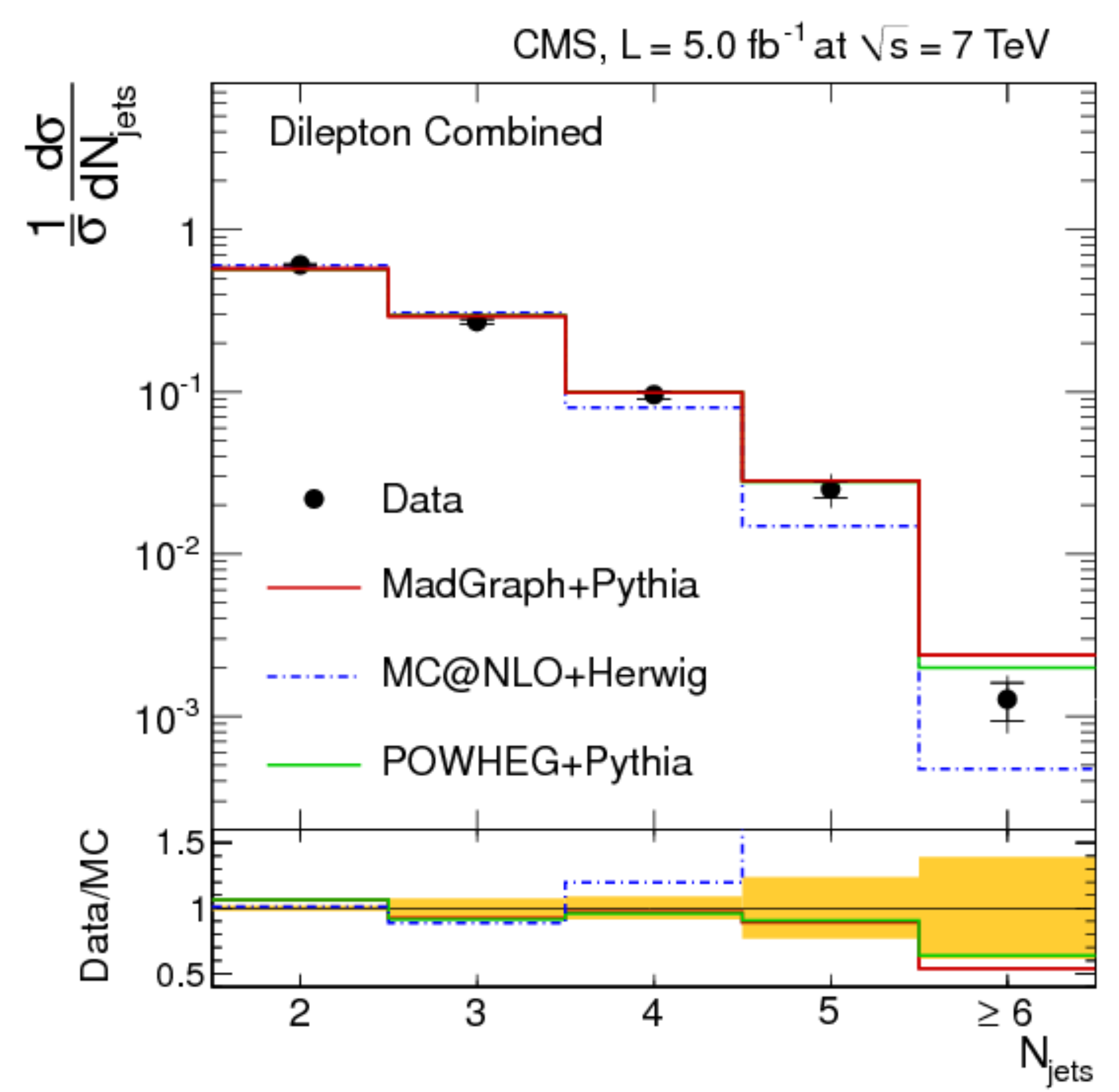}
\caption{(a) \ttjets\ cross section as a function of the jet multiplicity ($\ptj > 25\gev$)  measured by ATLAS in the lepton+jets channel~\cite{atltjets11}. (b)  \ttjets\  cross section as a function of the jet multiplicity ($\ptj > 30\gev$)  in the dilepton channel normalized to the inclusive \ttbar\ cross section  measured by the CMS experiment~\cite{cmstjets11}. \label{f:Vjets6}}
\end{figure}

\section{Summary and Outlook}

The LHC experiments ATLAS and CMS have measured  \Vjets\ and \ttjets\ final states over a large energy range  in data collected between 2010 and 2012 at $\rts = 7\tev$ and $\rts = 8\tev$. The results have been compared to  pQCD calculations at NLO and have been used to validate novel Monte Carlo techniques. In general, a good performance of  fixed-order NLO  calculations  and \meps\  generators is observed.  Some generators show tensions in more extreme phase space regions. 
Double and triple-differential measurements of   angles, invariant masses and transverse momenta can be used to constrain PDFs.
Many measurements at  $\rts = 8\tev$ are still ongoing and more results are expected in the coming year.




\appendix


\begin{thebibliography}{00}    



\bibitem{blackhat}
{C.~F.~Berger {\it et al.}},
\emph{Phys. Rev.}  {\bfseries D 78},  036003 (2008).


\bibitem{alpgen}
{M.~Mangano {\it et al.}}, 
\emph{JHEP} {\bfseries 0307}, 001 (2003).


\bibitem{sherpa}
{T.~Gleisberg {\it et al.}}, 
\emph{JHEP} {\bfseries 0902}, 007 (2009).

\bibitem{madgraph}
{J.~Allwall {\it et al.}},
\emph{JHEP} {\bfseries 06} (2011) 128.


\bibitem{sherpav2} 
{S.~H\"oche  {\it et al.}},
\emph{JHEP} {\bfseries 1304}, 027 (2013).


\bibitem{svd}
{A.~Hoecker, V.~Kartvelishvili},
\emph{Nucl. Instrum. Meth. } {\bfseries A 372}, 469 (1996).

\bibitem{dagostini}
{G.~D'Agostini}, 
\emph{Nucl. Instrum. Meth. } {\bfseries A 362}, 487 (1995).


\bibitem{antikt}
{M.~Cacciari, G.~P.~Salam and G.~Soyez},
5\emph{The \antikt\ jet clustering algorithm}, 
\emph{JHEP} {\bfseries 0804}, 063 (2008).



\bibitem{cdfzjets}
{CDF Collab. (T.~Aaltonen {\it et al.})}, 
\emph{Phys. Rev. Lett.} {\bfseries 100}, 102001 (2008).
\bibitem{cdfwjets}
{CDF Collab. (T.~Aaltonen {\it et al.})}, 
\emph{Phys. Rev. } {\bfseries D 77}, 011108 (2008).
\bibitem{d0zjetsa}
{D0 Collab. (V.~M.~Abazov {\it et al.})},
\emph{Phys. Lett.,}  {\bfseries B 669}, 278  (2008).
\bibitem{d0zjetsb}
{D0 Collab. (V.~M.~Abazov {\it et al.})},
\emph{Phys. Lett.} {\bfseries B 682}, 370  (2010).  


\bibitem{z2jetsnlo}
{J.~Campbell and R.~K.~Ellis},
\emph{Phys.Rev.} {\bfseries D65}, 113007 (2002).

\bibitem{bhwjets}
{C.~F.~Berger {\it et al.}},
\emph{Phys. Rev.}  {\bfseries D 80},  074036 (2009);
{C. F. Berger {\it et al.}}, 
\emph{Phys. Rev. Lett.} {\bfseries 106}, 092001 (2011).

\bibitem{bhzjets}
{C.~F.~Berger {\it et al.}},
\emph{Phys. Rev.}  {\bfseries D 82}, 074002  (2010);
{H.~Ita {\it et al.}}, 
\emph{Phys. Rev.} {\bfseries D 85} (2012) 031501;
{\it Z.~Bern {\it et al.}}, 
\emph{Phys. Rev.} {\bfseries D 88}, 014025 (2013).


\bibitem{atlzjets11}
{ATLAS Collab., (G.~Aad {\it et al.})},
\emph{JHEP} {\bfseries 07}, 032 (2013).

\bibitem{atlwjets11}
{ATLAS Collab., (G.~Aad {\it et al.})},
\emph{Eur.Phys.J.} {\bfseries C 75}, 75:82 (2015).

\bibitem{atlrjets11}
{ATLAS Collab., (G.~Aad {\it et al.})},
\emph{Eur.Phys.J.} {\bfseries C 74}, 3168 (2014). 


\bibitem{atlzjets12}
{ATLAS Collab., (G.~Aad {\it et al.})}, 
\emph{JHEP} {\bfseries 04}, 031 (2014).


\bibitem{cmszjets11}
{CMS Collab. (S.~Chatrchyan {\it et al.})},
\emph{Phys. Rev.} {\bfseries D 91}, 052008 (2015);
{CMS Collab. (S.~Chatrchyan {\it et al.})},
\emph{Phys. Lett.} {\bfseries B 722}, 238 (2013);
{CMS Collab. (S.~Chatrchyan {\it et al.})},
\emph{Phys. Rev.} {\bfseries D 88}, 112009 (2013).


\bibitem{cmswjets11}
{CMS Collab. (S.~Chatrchyan {\it et al.})},
\emph{Phys. Lett.} {\bfseries B 741}, 12 (2015).


\bibitem{powheg}
{S.~Alioli {\it et al.}},
\emph{JHEP} {\bfseries 1006},  043 (2010).




\bibitem{atlgjets10a}
{ATLAS Collab., (G.~Aad {\it et al.})}, 
\emph{Phys. Rev.} {\bfseries D 85}, 092014 (2012).


\bibitem{atlgjets10b}
{ATLAS Collab., (G.~Aad {\it et al.})}, 
\emph{Nucl. Phys.}  {\bfseries B 875}, 483-535 (2013). 


\bibitem{cmsgjets11}
{CMS Collab. (S.~Chatrchyan {\it et al.})},
\emph{JHEP} {\bfseries 06}, 009 (2014).

\bibitem{jetphox}
{S.~Catani  {\it et al.}},
\emph{JHEP} {\bfseries 05}, 028 (2002). 

\bibitem{photjetpdf}
{L.~Carminati {\it et al.}},
\emph{Europhys. Lett.}  {\bfseries 101}, 61002 (2013).


\bibitem{atltjets11}
{ATLAS Collab., (G.~Aad {\it et al.})}, 
\emph{JHEP} {\bfseries 01}, 020 (2015).

\bibitem{cmstjets11}
{CMS Collab. (S.~Chatrchyan {\it et al.})},
\emph{Eur.Phys.J.} {\bfseries C 74},  3014 (2014).


\end{thebibliography}
\end{document}